\documentclass[aps,prc,twocolumn,superscriptaddress,showpacs,floatfix]{revtex4}
\usepackage{mathrsfs}
\usepackage{latexsym}
\usepackage{amsmath}
\usepackage{amssymb}
\usepackage{graphicx}
\usepackage{longtable}

\usepackage{bbm}
\usepackage{epsfig}
\usepackage[usenames]{color}
\usepackage[colorlinks=true,citecolor=blue,linkcolor=blue]{hyperref}
\usepackage{colortbl}

\begin{document}



\title{Hadron-quark phase transition in asymmetric matter with dynamical
quark masses}

\author{G.Y. Shao}
\affiliation{INFN-Laboratori Nazionali del Sud, Via S. Sofia 62, I-95123
Catania, Italy}

\author{M. Di Toro}
\email[Corresponding author: ]{ditoro@lns.infn.it}
\affiliation{INFN-Laboratori Nazionali del Sud, Via S. Sofia 62, I-95123
Catania, Italy}
\affiliation{Physics and Astronomy Dept., University of Catania, Italy}

\author{B. Liu}
\affiliation{IHEP, Chinese Academy of Sciences, Beijing, China}
\affiliation{Theoretical Physics Center for Scientific Facilities, \\Chinese
Academy of Sciences, Beijing, China}

\author{M. Colonna}
\affiliation{INFN-Laboratori Nazionali del Sud, Via S. Sofia 62, I-95123
Catania, Italy}

\author{V. Greco}
\affiliation{INFN-Laboratori Nazionali del Sud, Via S. Sofia 62, I-95123
Catania, Italy}
\affiliation{Physics and Astronomy Dept., University of Catania}

\author{Y.X. Liu}
\affiliation{Department of Physics and State Key Laboratory of \\
Nuclear Physics and Technology,
Peking University, Beijing 100871, China}
\affiliation{Center of Theoretical Nuclear Physics,\\ National Laboratory of
Heavy Ion Accelerator, Lanzhou 730000, China}

\author{S. Plumari}
\affiliation{INFN-Laboratori Nazionali del Sud, Via S. Sofia 62, I-95123
Catania, Italy}
\affiliation{Physics and Astronomy Dept., University of Catania}


\begin{abstract}
The two-Equation of State (EoS) model is used to describe the hadron-quark
phase transition in asymmetric matter formed at high density in heavy-ion
collisions.
For the quark phase, the three-flavor Nambu--Jona-Lasinio~(NJL) effective
theory is used to
investigate the influence
of dynamical quark mass effects on the phase transition. 
At variance to the MIT-Bag results, with fixed current quark masses,
 the main important effect of the chiral dynamics is the appearance of an 
End-Point for the coexistence zone.
We show that a first order hadron-quark phase transition
may take place in the region T\,$\subset(50-80)$\,MeV and
$\rho_B^{}\subset(2-4)\,\rho_0$, which is possible to
be probed in the new planned facilities, such as FAIR at GSI-Darmstadt
and NICA at JINR-Dubna. From isospin properties of the mixed phase some
possible signals are suggested. The importance of chiral symmetry
and dynamical quark mass on the hadron-quark phase transition is stressed.
The difficulty of an exact location of Critical-End-Point comes from
its appearance in a region of competition between chiral symmetry breaking 
and confinement, where our knowledge of effective QCD theories is still
rather uncertain.   
\end{abstract}


\maketitle

\section{Introduction}
The determination of the phase diagram of strongly interacting matter and
the search for signals of the hadron-quark phase transition are challenges 
both in
theory and experiment.
Intensive studies on these fields have been developed in the last decades
\cite{Brown90,Stephanov98,Halasz98,Fodor02,Kawamoto07,
Toublan03,Werth05,Abuki06,Weise07,Andronic10,
Wambach09,Buballa09,Schaefer10,Herbst11,Fukushima11}.
Most phase diagrams have been derived from  Monte Carlo calculations of Lattice
QCD~\cite{Brown90,Fodor02,Kawamoto07} or
effective chiral  models~\cite{Toublan03,Werth05,
Abuki06,Fukushima11} with quark degree of freedom.
In the chiral effective model, one can describe well the line of chiral phase
transition and  the complicated
phase diagram of color superconductivity~\cite{
Shovkovy03,Huang03,Alford08}, while the
confinement phase transition
for the gluon part can be investigated with the extended NJL model coupled to
 the Polyakov loop
~\cite{Weise07,Buballa09,Schaefer10,Herbst11,Fukushima04,
Kashiwa08,Abuki08,Fu08}) 
in the temperature
and chemical potential plane. However a large uncertainty remains about
the derivation of baryon density and 
energy density
of the onset of the hadron-quark phase transition~\cite{Herbst11,Stephanov06}.

Although there are attempts to describe nuclear and quark matter in
unified effective
models~\cite{Lawley06,Rezaeian06,Bentz06,Dexheimer10,Stoecker11}, 
further investigations are needed to give more satisfying results,
 in particular at high baryon and isospin densities, of interest for the
expected phase transition in heavy ion collisions and compact stars.


An alternative approach to describe the hadron-quark phase transition is 
based on
a two-EoS model with the Gibbs criteria, which has been widely used to make
predictions on the phase transition
in the interior of neutron stars~(e.g.,\,\cite{Glendenning92,Glendenning98,
Burgio02,Maruyama07,Yang08,Shao10,Xu10} ).

We remark that, even in the two-EoS approach, only a few papers have studied 
the phase diagram of
hadron-quark transitions at high baryon density in connection to the 
phenomenology of
heavy-ion collision ten A GeV range (intermediate energies)
~\cite{Muller97, Toro06, Torohq09, Pagliara10, Cavagnoli10}.
In Ref.~\cite{Muller97} the phase transition
from hadron to quark matter has been firstly analyzed for isospin asymmetric
matter. It should be noticed that recently increasing  attention is paid to 
this aspect, and some
observable effects
are suggested to be seen in charged meson yield ratio and in the
onset of quark number scaling of the meson/baryon elliptic flows 
\cite{Toro06, Torohq09}. 
This provides us a new orientation to
investigate the hadron-quark
phase transition, and can stimulate some new relevant research in the field.
Later, hyperons have been included in \cite{Cavagnoli10}, but the calculated
results show that strange
baryons are not important due to the exiguous final population in the short
time scale of a nucleus-nucleus collision.
Furthermore, the Cooper pair effect of $u,\,d$ quarks
(two flavor superconductor, 2SC) has been considered in the quark
phase~\cite{Pagliara10}, which reduces the symmetry energy
difference between hadronic and quark phases. The most important conclusion
emerging from these works
is that the onset density of hadron-quark phase transition is smaller in
asymmetric matter
than that in symmetric matter, which is possibly reached through
heavy-ion reactions at new planned facilities, such as FAIR at
GSI-Darmstadt and NICA at JINR-Dubna, where heavy ion beams~(even unstable,
with large isospin asymmetry) will be available with good intensities in the 
1-30 A GeV energy region.


One drawback in all these calculations is that current mass (or massless)
$u,\,d$ quarks are taken for the quark phase, where the MIT-Bag like models
are used. The obtained results are possibly reliable at
high density, after the restoration of chiral symmetry due to the asymptotic
freedom of QCD, but
the chiral symmetry breaking at finite temperature and
low densities is not accounted for.
From $\rho\sim 4\,\rho_0$ to 
smaller densities, the dynamical masses of $u,\,d$ quarks become larger and
larger, and they almost reach
 $200\,$MeV at $\rho \sim \rho_0$ for some parameter sets. This means that 
non-perturbative effects become more and more important
with the decreasing baryon density.
Therefore, calculations of properties of the phase diagram
at low density and finite temperature are not fully consistent. As a matter 
of fact when the MIT-Bag
model is taken for the quark part, we see that
the T-$\rho$, T-$\mu$ phase diagrams highly depend on the values of
the bag constant $B$, which cannot be determined accurately~\cite{Torohq09}.

In order to obtain more reliable theoretical results and  predict possible
observables in the planned experiments,
we take the Nambu-Jona Lasinio (NJL) model to describe the quark phase with
the interaction between quarks,
where the chiral symmetry breaking and restoration are well described.

It was also proposed in Ref.~\cite{Greiner87}, that in high energy
heavy-ion collisions strange and antistrange quarks can be produced by thermal
excitation
(with net strangeness being zero required by the conservation law of
strangeness in strong interaction)
and strangeness would be much more abundant in the quark component.
Therefore we will also consider
the thermal excitation of strange and antistrange quark, but we need to
keep the chemical potential of strange quark to be zero, $\mu_s=0$, before
the beginning of hadronization
in the expanding process to make sure of the net strangeness being zero.
One mechanism of hadron production is quark
recombination~\cite{Srivastava95,Biro95,Hwa03,Fries03,Greco03,Fries08},
which is out of the range of the discussion in this paper.

The paper is organized as follows. In Section II, we describe
briefly the used effective Lagrangians and give the relevant formulas for
the Relativistic Mean Field ($RMF$) theory adopted for the hadron sector. In
Section III, we present the calculated
phase diagrams and compare them with those obtained in the MIT-Bag model.
Besides, we present some discussions and conclusions.
Finally, a summary is given in Section IV.

\section{ nuclear matter, quark matter and the mixed phase}
For the hadron phase, the non-linear Relativistic Mean Field (RMF) approach is
used, which
provides an excellent description of  nuclear matter and finite nuclei. One can
calibrate the hadronic equation of state at zero temperature and normal
nuclear densities,
and then extrapolate into the regime of finite density and temperature.
Our parametrizations are also tuned to reproduce collective flows and
particle production at higher energies, where some hot and dense matter
is probed, see~\cite{Toro09} and refs. therein.
The exchanged mesons include the isoscalar-scalar
meson($\sigma$), isoscalar-vector meson($\omega$), isovector-vector
meson($\rho$) and  isovector-scalar meson($\delta$). The effective Lagrangian
can be written as
\begin{widetext}
\begin{eqnarray}
\cal{L} &=&\bar{\psi}[i\gamma_{\mu}\partial^{\mu}- M
          +g_{\sigma }\sigma+g_{\delta }\boldsymbol\tau_{}
\cdot\boldsymbol\delta
          -g_{\omega }\gamma_{\mu}\omega^{\mu}
          -g_{\rho }\gamma_{\mu}\boldsymbol\tau_{}\cdot\boldsymbol
\rho^{\mu}]\psi
           \nonumber\\
   & &{}+\frac{1}{2}\left(\partial_{\mu}\sigma\partial^
{\mu}\sigma-m_{\sigma}^{2}\sigma^{2}\right)
       - \frac{1}{3} b\,(g_{\sigma} \sigma)^3-\frac{1}{4} c\,
(g_{\sigma} \sigma)^4
          +\frac{1}{2}\left(\partial_{\mu}\delta\partial^{\mu}\delta
          -m_{\delta}^{2}\delta^{2}\right)  \nonumber\\
       & &{}+\frac{1}{2}m^{2}_{\omega} \omega_{\mu}\omega^{\mu}
          -\frac{1}{4}\omega_{\mu\nu}\omega^{\mu\nu}
          +\frac{1}{2}m^{2}_{\rho}\boldsymbol\rho_{\mu}\cdot\boldsymbol
\rho^{\mu}
          -\frac{1}{4}\boldsymbol\rho_{\mu\nu}\cdot\boldsymbol\rho^{\mu\nu},
 \end{eqnarray}
\end{widetext}
where the antisymmetric tensors of vector mesons are given by
\begin{equation}
\omega_{\mu\nu}= \partial_\mu \omega_\nu - \partial_\nu
\omega_\mu,\qquad \nonumber \rho_{\mu\nu} \equiv\partial_\mu
\boldsymbol\rho_\nu -\partial_\nu \boldsymbol\rho_\mu.
\end{equation}

In the $RMF$ approach all effective meson fields can be expressed via
their mean values, simply related to baryon and scalar nucleon densities.
In this way only nucleon degress of freedom are left to describe dynamics and 
thermodynamics of the system \cite{Liu02,Baran05}.
The nucleon chemical potential and effective mass in nuclear medium are
\begin{equation}
\mu_{i}^{} =\mu_{i}^{*}+g_{\omega
}\omega+g_{\rho}\tau_{3i}^{}\rho \, ,
\end{equation}
\begin{equation}
M_{i}^{*} = M -g_{\sigma }^{}\sigma-g_{\delta
}^{} \tau_{3i}^{} \delta,
\end{equation}
where $M$ is the free nucleon mass, $ \tau_{3p}^{}=1$ for proton and 
$\tau_{3n}^{}=-1$ for neutron, and $\mu_{i}^{*}$
is the effective chemical potential which reduces to Fermi energy
$E_{Fi}^{*}=\sqrt{k_{F}^{i^{2}}+M_{i}^{*^{2}}}$ at zero temperature.
The baryon and isospin chemical potentials in the hadron phase are defined as
\begin{equation}
\mu_{B}^{H} =\frac{\mu_{p}+\mu_{n}}{2},\ \ \ \ \ \mu_{3}^{H} =\frac
{\mu_{p}-\mu_{n}}{2}.
\end{equation}

The energy density and pressure of nuclear matter at finite temperature can
be derived as
\begin{widetext}
\begin{equation}
\varepsilon^{H} =  \sum_{i=p,n}\frac{2}{(2\pi)^3} \int \! d^3
\boldsymbol k \sqrt{k^2 + {M^*_i}^2}(f_{i}(k)+\bar{f}_{i}(k))+
\frac{1}{2}m_\sigma^2 \sigma^2 +
\frac{b}{3}\,(g_{\sigma }^{} \sigma)^3+ \frac{c}{4}\,(g_{\sigma
}^{} \sigma)^4
+\frac{1}{2}m_\delta^2 \delta^2+\frac{1}{2}m_\omega^2 \omega^2 +
\frac{1}{2}m_\rho^2 \rho^2 \, ,
\end{equation}
\begin{equation}
P^{H}  =  \sum_{i=p,n} \frac{1}{3}\frac{2}{(2\pi)^3} \int \!
d^3 \boldsymbol k \frac{k^2}{\sqrt{k^2 + {M^*_i}^2}}(f_{i}(k)+
\bar{f}_{i}(k))- \frac{1}{2}m_\sigma^2
\sigma^2 - \frac{b}{3}\,(g_{\sigma }^{} \sigma)^3 -
\frac{c}{4}\,(g_{\sigma }^{} \sigma)^4
-\frac{1}{2}m_\delta^2 \delta^2+\frac{1}{2}m_\omega^2 \omega^2 +
\frac{1}{2}m_\rho^2 \rho^2 \, .
\end{equation}
\end{widetext}
where $f_{i}(k)$ and $\bar{f}_{i}(k)$ are the fermion and antifermion
distribution functions
for protons and neutrons ($i=p,\,n$):
\begin{equation}
  f_{i}(k)=\frac{1}{1+\texttt{exp}\{(E_{i}^{*}(k)-\mu_{i}^{*})/T\}} ,
\end{equation}
\begin{equation}
  \bar{f}_{i}(k)=\frac{1}{1+\texttt{exp}\{(E_{i}^{*}(k)+\mu_{i}^{*})/T\}}.
\end{equation}
The effective chemical potentials $\mu_{i}^{*}$ are determined by the nucleon
densities
\begin{equation}\label{hadrondensity}
  \rho_i= 2 \int \! \frac{d^3 \boldsymbol k}{(2\pi)^{3}}( f_{i}(k)-
\bar{f}_{i}(k)).
\end{equation}
With the baryon density $\rho=\rho_{B}^{H}=\rho_p+\rho_n$ and isospin
density $\rho_{3}^{H}=\rho_p-\rho_n$,
the asymmetric parameter can be defined as
\begin{equation}
   \alpha^{H}\equiv-\frac{\rho_{3}^{H}}{\rho_{B}^{H}}=\frac{\rho_p-
\rho_n}{\rho_p+\rho_n}.
\end{equation}
As model Lagrangians for the hadron phase,
$NL\rho$~(the isovector scalar meson
$\delta$ being not included)
and $NL\rho\delta$~(with the $\delta$ meson) will be used. The effective
meson couplings heve been chosen to reproduce good  nuclear matter properties
and even to
represent
a reasonable average of the density dependence predicted by
Relativistic Dirac-Brueckner-Hartree-Fock
calculation~\cite{Hofmann01,Goegelein08}, See details in
Appendix A1 of Ref.~\cite{Toro06, Torohq09}
and also Refs.~\cite{Liu02,Baran05,Toro09}. A note about strangeness
in the hadron phase can be found in Ref.~\cite{hstrange}.

For quark matter, we use the NJL model~\cite{Nambu61} to describe the
interaction
between quarks which is responsible for the quark dynamics at intermediate
energies.
The NJL model describes well the mesons spectra and successfully explains
the dynamics of spontaneous
breaking of chiral symmetry and its restoration at high
densities/\,chemical potential~\cite{Volkov84,Hatsuda84,Klevansky92,Hatsuda94,
Alkofer96,Buballa05}.
The Lagrangian density in the three-flavor NJL model is taken as
\begin{eqnarray}
\mathcal{L}_{q}&=&\bar{q}(i\gamma^{\mu}\partial_{\mu}-\hat{m}_{0})q+
G\sum_{k=0}^{8}[(\bar{q}\lambda_{k}q)^{2}+
(\bar{q}i\gamma_{5}\lambda_{k}q)^{2}]\nonumber \\
           &&-K[\texttt{det}_{f}(\bar{q}(1+\gamma_{5})q)+\texttt{det}_{f}
(\bar{q}(1-\gamma_{5})q)].
\end{eqnarray}
where $q$ denotes the quark fields with three flavors, $u,\ d$, and
$s$, and three colors; $\hat{m}_{0}=\texttt{diag}(m_{u},\ m_{d},\
m_{s})$ in flavor space; $\lambda_{k}$ are the Gell-Mann matrices and
 $G$ and $K$  the four-point and
six-point interacting constants, respectively. The four-point
interaction term in the Lagrangian keeps the $SU_{V}(3)\times
SU_{A}(3)\times U_{V}(1)\times U_{A}(1)$ symmetry, while the 't
Hooft six-point interaction term breaks the $U_{A}(1)$ symmetry.

As an effective model, the NJL model is not
renormalizeable, so a cut-off $\Lambda$ is implemented in 3-momentum
space for divergent integration. We take the model parameters:
$\Lambda=603.2$ MeV, $G\Lambda^{2}=1.835$, $K\Lambda^{5}=12.36$,
$m_{u,d}=5.5$  and $m_{s}=140.7$ MeV, determined by Rehberg, Klevansky,
and H\"{u}fner in Ref.~\cite{Rehberg95}
by fitting $f_{\pi},\ M_{\pi},\ m_{K}$ and $\ m_{\eta}$ to their
experimental values.

The thermodynamical potential in 3-flavor quark system in mean field
approximation is
\begin{widetext}
\begin{eqnarray}
\Omega^{Q}&\!\!\!\!\!\!=\!\!\!\!\!\!&-2N_{c}\!\!\sum_{i=u,d,s}\int\frac
{d^{3}\boldsymbol k}{(2\pi)^{3}}\bigg\{
\beta^{-1}\ln\Big[1+e^{-\beta(E_{i}(k)-\mu_{i})}\Big]\!\!+
\beta^{-1}\ln\Big[1+e^{-\beta(E_{i}(k)+\mu_{i})}\Big]\,
\nonumber\\
&&{}+E_{i}\bigg\}+2G\left({\phi_{u}}^{2}
+{\phi_{d}}^{2}+{\phi_{s}}^{2}\right)-4K\phi_{u}\,\phi_{d}\,\phi_{s}+C\,,
\label{therm-pot}
\end{eqnarray}
\end{widetext}
where $C$ is a constant to be fixed by physics conditions, $N_c=3$ is 
the number of color degrees of freedom,
$E_{i}(k)=\sqrt{k^{2}+M_{i}^{*2}}$ is energy-momentum dispersion relation
of the quark flavor $i$, and $\mu_i$ is the corresponding chemical
potential. The dynamical quark masses and quark condensates are
coupled by the following equations:
\begin{equation}
M_{i}^*=m_{i}-4G\phi_i+2K\phi_j\phi_k\ \ \ \ \ \ (i\neq j\neq k),
\label{mass}
\end{equation}
\begin{equation}
\phi_i=-2N_{c}\int_{\Lambda}\frac{d^{3}\boldsymbol k}{(2\pi)^{3}}
\frac{M_{i}^*}{E_{i}}
\big(1-n_i(k)-\bar{n}_i(k)\big),
\end{equation}
here $n_i(k)$ and $\bar{n}_i(k)$ are Fermi-Dirac distribution function of
quark and antiquark
\begin{equation}
  n_{i}(k)=\frac{1}{1+\texttt{exp}\{(E_{i}^{}(k)-\mu_{i}^{})/T\}} ,
\end{equation}
and
\begin{equation}
  \bar{n}_{i}(k)=\frac{1}{1+\texttt{exp}\{(E_{i}^{}(k)+\mu_{i}^{})/T\}}.
\end{equation}

The pressure and energy density of the quark system are:
\begin{widetext}
\begin{eqnarray}
P^{Q}&=&{}2N_{c}\sum_{i=u,d,s}\int\frac{d^{3}\boldsymbol k}{(2\pi)^{3}}
\frac{k^2}{3E_{i}(k)}(n_i(k)+\bar{n}_i(k))+2N_{c}\sum_{i=u,d,s}
\int_{\Lambda}\frac{d^{3}\boldsymbol k}{(2\pi)^{3}}E_{i}(k)-2G
\left({\phi_{u}}^{2}
+{\phi_{d}}^{2}+{\phi_{s}}^{2}\right)\nonumber\\
&&{}+4K\phi_{u}\,\phi_{d}\,\phi_{s}\,, \label{QM_pressure}
\end{eqnarray}
\begin{eqnarray}
\varepsilon^{Q}&=&{}2N_{c}\sum_{i=u,d,s}\int\frac{d^{3}\boldsymbol k}
{(2\pi)^{3}}
E_{i}(k)(n_i(k)+\bar{n}_i(k))-2N_{c}\sum_{i=u,d,s}\int_{\Lambda}\frac{d^{3}
\boldsymbol k}{(2\pi)^{3}}E_{i}(k)+2G\left({\phi_{u}}^{2}
+{\phi_{d}}^{2}+{\phi_{s}}^{2}\right)\nonumber\\
&&{}-4K\phi_{u}\,\phi_{d}\,\phi_{s}\,. \label{QM_energy}
\end{eqnarray}
\end{widetext}
If we define
\begin{eqnarray}
B_{eff}&=&-2N_{c}\sum_{i=u,d,s}\int_{\Lambda}\frac{d^{3}\boldsymbol k}
{(2\pi)^{3}}E_{i}(k)\nonumber\\
&&{}+2G\left({\phi_{u}}^{2}
+{\phi_{d}}^{2}+{\phi_{s}}^{2}\right)-4K\phi_{u}\,\phi_{d}\,
\phi_{s}\,,\nonumber\\ \label{Beff}
\end{eqnarray}
we will find $B_{eff}$ acting exactly as an effective bag constant,
just like the bag constant in the MIT-Bag model. The difference  is
that now  $B_{eff}$ depends on the interaction between different
quarks as well as on density and temperature. The same holds true
for the Fermi motion contribution. For all the integrations the cut-off 
is implemented, together with the irrelevant thermodynamical constant $C$
of Eq.\ref{therm-pot}, as usual by requiring
the energy density and pressure equal to zero in the vacuum,
\emph{ie.}, $\varepsilon^{Q}=P^{Q}=0$ at zero temperature and
density.

The number density of quark flavor $i$ can be derived with the relation
$\rho_{i}=-\partial \Omega_{q}/\partial \mu_{i}$
\begin{equation}
\rho_{i}=2N_{c}\int\frac{d^{3}\boldsymbol k}{(2\pi)^{3}}(n_{i}(k)-
\bar{n}_{i}(k))\,.
\label{quarkdensity}
\end{equation}
As we have mentioned in the Introduction, although the strangeness could be
produced at finite temperature  in
heavy-ion collision, the number of net strange quark is zero before the
hadronization takes place in
the expanding process. So we must keep $\mu_{s}=0$ in the calculation
according to Eq.~\eqref{quarkdensity}, but the strange quarks in loop
diagrams do contribute to the $u,\,d$ self energies
via a six-point interaction, and then they can affect the dynamical mass
of $u,\,d$ quark according to Eq.~\eqref{mass}.
The corresponding definitions of density and chemical potential in quark
phase are as follows
\begin{equation}
\rho_{B}^Q=\frac{1}{3}(\rho_u+\rho_d),\ \ \ \ \rho_{3}^Q=\rho_u-\rho_d,
\end{equation}
\begin{equation}
\mu_{B}^Q=\frac{3}{2}(\mu_u+\mu_d),\ \ \ \ \mu_{3}^Q=\frac{1}{2}(\mu_u-\mu_d).
\end{equation}
The asymmetry parameter for quark phase is defined by
\begin{equation}
   \alpha^{Q}\equiv-\frac{\rho_{3}^{Q}}{\rho_{B}^{Q}}=3\frac{\rho_d-\rho_u}
{\rho_u+\rho_d}.
\end{equation}

The above introduction is a separate description of the hadronic and the quark
phase. The goal of the present work is
to extend these studies to the hadron-quark phase transition, in particular
to stress the effect of dynamical quark masses.

Since there are two conserved quantity, baryon number and
isospin, during the phase transition, the Gibbs criteria
(thermal, chemical and mechanical equilibrium) should be
implemented for the mixed phase. General discussion
of phase transitions in multicomponent systems can be found
in Ref.\cite{Glendenning92}.

If we define $\chi$ the fraction of quark matter in the
mixed phase, the Gibbs conditions for the mixed phase are
\begin{eqnarray}
& &\mu_B^H(\rho_B^{},\rho_3^{},T)=\mu_B^Q(\rho_B^{},\rho_3^{},T)\nonumber\\
& &\mu_3^H(\rho_B^{},\rho_3^{},T)=\mu_3^Q(\rho_B^{},\rho_3^{},T)\nonumber\\
& &P^H(\rho_B^{},\rho_3^{},T)=P^Q(\rho_B^{},\rho_3^{},T),
\end{eqnarray}
where $\rho_B^{}=(1-\chi)\rho_B^{H}+\chi \rho_B^{Q}$ is the total
baryon density in the mixed phase,
and the total isospin density is $\rho_3^{}=(1-\chi)\rho_3^{H}+\chi \rho_3^{Q}$.
With the initial condition of asymmetric parameter $\alpha^H$ in
heavy-ion collision, the global asymmetry parameter $\alpha$ for
the mixed phase should be
\begin{eqnarray}
& &   \alpha\equiv-\frac{\rho_{3}^{}}{\rho_{B}^{}}= -\frac{(1-\chi)\rho_3^{H}+
\chi \rho_3^{Q}}{(1-\chi)\rho_B^{H}+\chi \rho_B^{Q}}~,\nonumber\\
& & \alpha^H(\chi=0) =  \alpha^Q(\chi=1)
\end{eqnarray}
according to the charge conservation.

\section{NUMERICAL RESULTS AND DISCUSSION}
Before presenting the phase diagram of hadron-quark phase
transitions with the NJL quark model, we firstly display some
results with the MIT-Bag model in which free fermions are considered
with the current mass
 $m_u=m_d=5.5\,$MeV. As for  the bag constant, different values are used
 in literature, here
 we take $(B_{\texttt{MIT}})^{1/4}=160\,$MeV and 190\,MeV, respectively, for 
comparison.
All that will be useful for general comments on the nature of the phase 
transition as well as for a better understanding of the dynamical mass 
effects in the NJL approach.

We will pay more attention
to the phase diagram of asymmetric matter. Indeed
the onset density of the hadron-quark phase transition for
asymmetric matter is smaller than in symmetric matter
 \cite{Muller97,Toro06,Torohq09,Cavagnoli10,Pagliara10}, then with the 
possibility to probe it in
heavy-ion collision at the new planned facilities, for example
FAIR at GSI-Darmstadt and NICA at JINR-Dubna,  We focus the attention
on experiments with neutron-rich stable heavy beams, where large
intensities can be reached. Here the largest accessible isospin
asymmetry parameter $\alpha$ is just above 0.2 (e.g. we have
$\alpha=0.227$ in $^{238}$U+$^{238}$U collision, see also the Table
II in Ref.\cite{Cavagnoli10}). Therefore we will consider
$\alpha=0.2$ in our calculation, just as taken in
Refs.~\cite{Toro06, Torohq09, Pagliara10, Cavagnoli10}. Of course
the use of more asymmetric unstable beams will enhance the isospin
effects described here.

\subsection{Some results with the MIT-Bag quark EoS}

\begin{figure}[htbp]
\begin{center}
\includegraphics[scale=0.25]{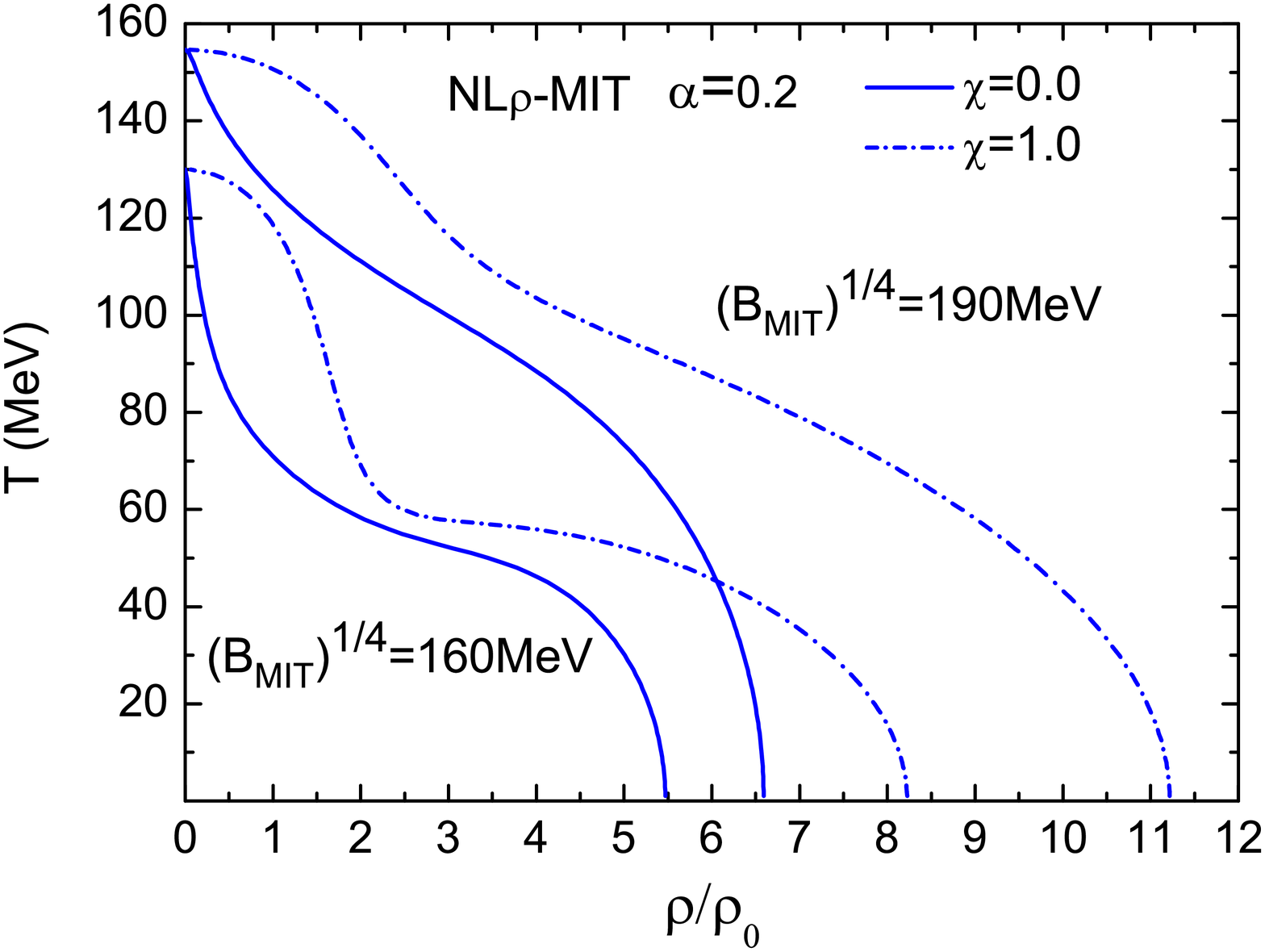}
\caption{\label{fig:nlrhomit-Trhoasym}(Color on line)~
The $T-\rho_B$ plane of
asymmetric matter with the isospin ratio  $\alpha=0.2$ for
$(B_{\texttt{MIT}})^{1/4}=160\,$MeV and 190\,MeV. The region between
the solid and dash-dot curve gives the binodal surface of the mixed
phase, where the hadron and quark matter coexist. $NL\rho$
parametrization is used for the hadron phase.}
\end{center}
\end{figure}

\begin{figure}[htbp]
\begin{center}
\includegraphics[scale=0.25]{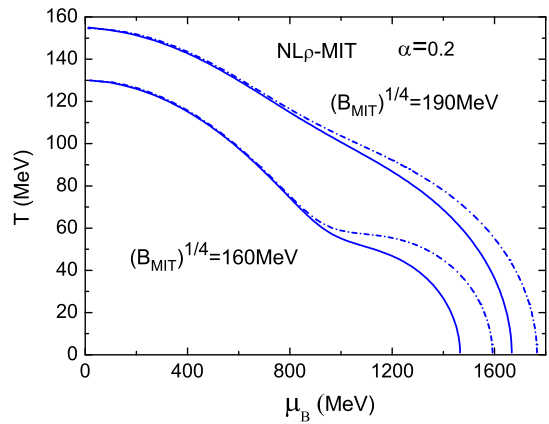}
\caption{\label{fig:nlrhomit-Tmuasym}(Color on line)~
The $T-\mu_B$ plane of
asymmetric matter with the isospin ratio  $\alpha=0.2$  for
$(B_{\texttt{MIT}})^{1/4}=160\,$MeV and 190\,MeV. The curves are
like in the previous figure. $NL\rho$ parametrization is used for
the hadron phase.}
\end{center}
\end{figure}

In Figs.~\ref{fig:nlrhomit-Trhoasym} and~\ref{fig:nlrhomit-Tmuasym}
we plot the $T-\rho_B$ and $T-\mu_B^{}$ phase diagrams . In each
figure the solid curves are the phase-transition lines from nuclear
matter to quark matter for $(B_{\texttt{MIT}})^{1/4}=160\,$ MeV
(190\,~MeV), and the corresponding dash-dot curves are the transition
lines to pure quark matter. $NL\rho$ hadron $EoS$ has been used in
the calculation.

From the two figures, it easy to see that the decrease of bag
constant reduces the onset density of quark matter, and in general
that the phase transition curve highly depends on the value of bag
constant, which is one of the motivations of this paper to use the
NJL model to investigate the phase transition.
In the Fig.~\ref{fig:nlrhomit-Tmuasym} we see a variation of the
baryon chemical potential along the transition from pure hadron
to pure quark phase, that could be
interpreted as an evidence of a continuous second order phase
transition for asymmetric matter, see~\cite{Muller97}. We like to
comment this point that we will see also clearly in the following
results with the $NJL$ quark $EoS$.

The transition is of first order, with the presence of a coexistence
mixed phase, in both cases of symmetric and asymmetric matter. For
symmetric matter, with only one conserved charge $\rho_B$, we have
the expected discontinuities of thermodynamical quantities like
entropy, pressure, at fixed chemical potential $\mu_B$. The presence
of two conserved charges, $\rho_B$ and $\rho_3$, in asymmetric
matter keeps the first order nature of the transition but changes
some properties inside the mixed phase. Due to the presence of a new
degree of freedom the interaction can choose the most energetically
favored charge densities in each phase, at each relative
concentration, in order to minimize the free energy of the system
~\cite{Glendenning92}. We can have an increase of pressure and
chemical potential inside the mixed phase but the transition is
still of first order since for each $\chi$ fraction we have a
discontinuity in the $\rho_B$ and $\rho_3$ densities of the two
phases in equilibrium. We note that this effect, directly related to
the internal interaction in the two phases, leads to the important
isospin distillation mechanism ($\alpha^Q\,>\,\alpha^H$) inside the
mixed phase, as shown later, of interest for possible experimental
signals. Similar properties are present in the liquid-gas
transition for dilute asymmetric nuclear matter, which is behind the
multifragmentation processes at the Fermi energies \cite{Baran05,Col04}.

Before closing the discussion of MIT-Bag results we would like to add a few 
comments about the interpretation of the $T\,-\,\rho_B$ and $T\,-\,\mu_B$
phase diagrams of 
Figs.~\ref{fig:nlrhomit-Trhoasym},~\ref{fig:nlrhomit-Tmuasym}.

For the lower bag costant value ($B^{1/4}=160~MeV$) we clearly see a squeezing
of the binodal surface (mixed phase region) in the $T\,-\,\rho_B$ plane
at temperature $T\,\simeq\,60~MeV$, corresponding to a baryon chemical 
potential $\mu_B\,\simeq\,900~MeV$, and a re-opening at higher temperature 
and smaller 
density (or chemical potential). The same is happening for the large bag 
constant, although less evident.

We note that this effect is appearing just around a transition chemical 
potential  equal to the nucleon effective mass $\mu_B\,\simeq\,M_{n,p}^*$.
We have a simple interpretation:

\begin{itemize}

\item{ The ``opening'' at low baryon density and high temperature.

Here the transition $\mu_B$ is below the nucleon effective mass
(not much reduced at low densities) and the effective chemical potential
 $\mu_i^*$ even smaller. Thus in the hadron part 
 $E_{F,i}^*\,>\,\mu_i^*$ and the fermion distribution function of the nucleons
will be rather small.  Therefore $\rho_B^H$, Eq.~(\ref{hadrondensity}),
 will show a slow increase with
$\mu_B$, while the pressure, mostly of thermal nature, is increasing
 (even for the positive antifermion contribution).
At variance in the quark sector $\mu_B$ is always much larger
than the used current quark masses ($5.5~MeV$) and so we see a fast
 increase of $\rho_B^Q$ to balance the hadron pressure.}

\item{The ``re-opening'' at high baryon density and low temperature.

With the increasing of baryon density/chemical potential, the nucleon 
effective mass decreases and the transition $\mu_i\,<\,M_i^*$ to
 $\mu_i\,>\,M_i^*$ takes place. Thus the fermion distribution function
of nucleons and $\rho_B^H$ will show a fast 
increase with $\mu_B$. The interaction part of the hadron pressure will also
increase. In correspondence we will need a fast increase even of the 
$\rho_B^Q$ in order to keep the $P^H\,=\,P^Q$ balance.}

\end{itemize}

We remark that the first argument is not anymore valid in the NJL frame,
since the quark masses at low density and finite temperature recover the
much larger  constituent quark values. Therefore we would expect
important qualitative differences in this low $\mu_B$ phase diagram region
when we take into account the chiral mass dynamics, see Subsection C.

\subsection{Results with the $NJL$ quark $EoS$}

\begin{figure}[htbp]
\begin{center}
\includegraphics[scale=0.25]{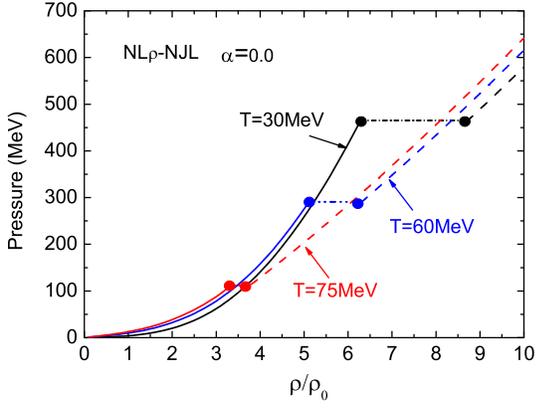}
\caption{\label{fig:nlrhonjl-Prhosym}(Color on line)~ 
Pressure of symmetric matter
as a function of baryon number density  at different temperatures.
The solid dots correspond to the limits of the mixed phase.
$NL\rho\,-\,NJL$ two-$EoS$ calculation.}
\end{center}
\end{figure}

\begin{figure}[htbp]
\begin{center}
\includegraphics[scale=0.25]{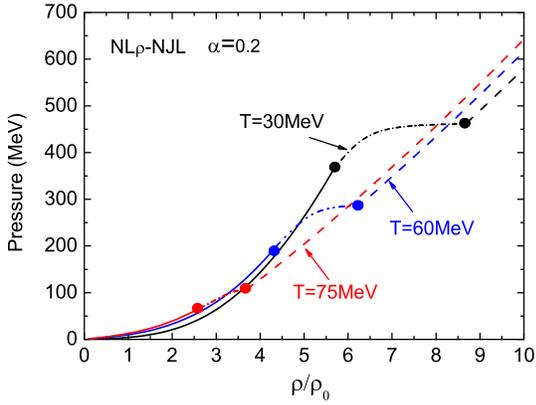}
\caption{\label{fig:nlrhonjl-Prhoasym}(Color on line)~ 
Like previous figure, but for
the asymmetric matter $\alpha=0.2$ case.}
\end{center}
\end{figure}

In Figs.~\ref{fig:nlrhonjl-Prhosym} and
Fig.~\ref{fig:nlrhonjl-Prhoasym} we present the $P- \rho_B$ phase
diagram, within the $NL\rho\,-\,NJL$ two-$EoS$ scheme,
 respectively for symmetric and asymmetric ($\alpha=0.2$) matter.
 For each temperature the mixed phase region is between the two solid
 dots.

Clearly the pressure is a constant in the mixed phase of symmetric
matter, Fig.~\ref{fig:nlrhonjl-Prhosym}, just the same as with
Maxwell construction. At variance in the asymmetric case,
Fig.~\ref{fig:nlrhonjl-Prhoasym}, we have a monotonous increase. It
is interesting to note that pressure rising is faster in the first
part of the mixed phase (more evident for the lower temperatures
T=30, 60 MeV). This is due to the isospin distillation effect, i.e.
a large $\rho_d\,-\,\rho_u$ asymmetry for reduced quark fractions,
see next Subsection, which is increasing the quark pressure as we
can expect from Eq.~\eqref{QM_pressure}.

 From Figs.~\ref{fig:nlrhonjl-Prhosym},~\ref{fig:nlrhonjl-Prhoasym},
  we can also see that the size of the
mixed phase shrinks with temperature. Meanwhile the onset density
becomes smaller, opening the possibility of probing the coexistence
phase in heavy-ion collision at intermediate energies. This effect
is further enhanced by the isospin asymmetry as we will discuss in
the next Subsection.


\subsubsection{Symmetry energy effects}
In order to study symmetry energy effects we have performed the
hadron-quark transition calculations with the hadron EoSs $NL\rho$
and $NL\rho\delta$ which present rather different symmetry terms at
high baryon density, stiffer when the $\delta$ meson is included 
\cite{Liu02,Baran05}.

\begin{figure}[htbp]
\begin{center}
\includegraphics[scale=0.25]{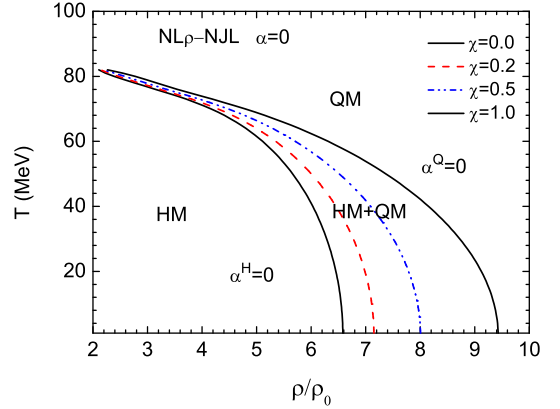}
\caption{\label{fig:nlnjl-Trhosym}(Color on line)~
Phase diagram in the $T-\rho_B^{}$
plane for symmetric matter with the parameter set $NL\rho$. Of
course the same curves are obtained with the parameter set $NL\rho
\delta$. }
\end{center}
\end{figure}

\begin{figure}[htbp]
\begin{center}
\includegraphics[scale=0.25]{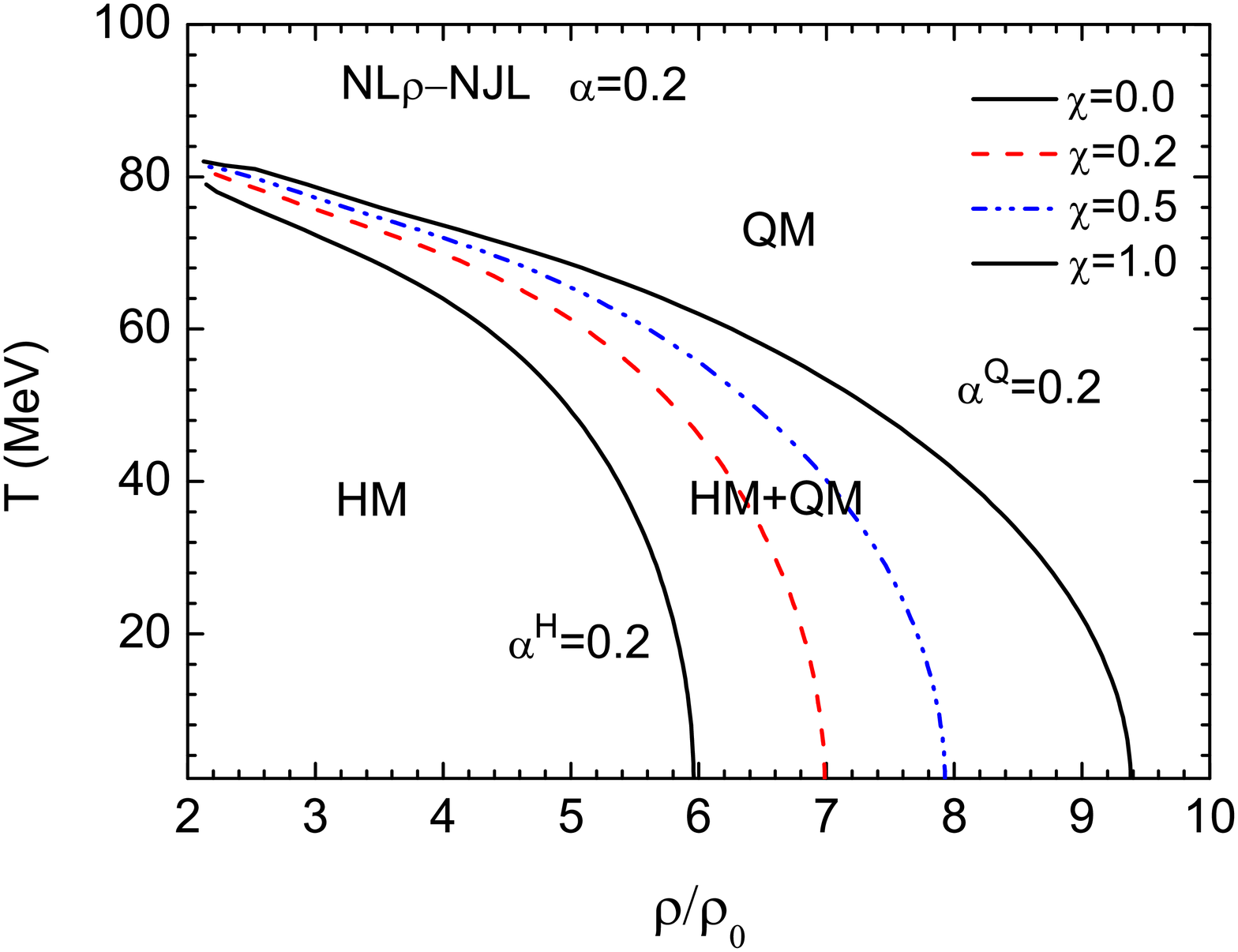}
\caption{\label{fig:nlrhonjl-Trhoasym}(Color on line)~
Phase diagram in the
$T-\rho_B^{}$ plane for asymmetric matter with the parameter set
$NL\rho$.}
\end{center}
\end{figure}

\begin{figure}[htbp]
\begin{center}
\includegraphics[scale=0.25]{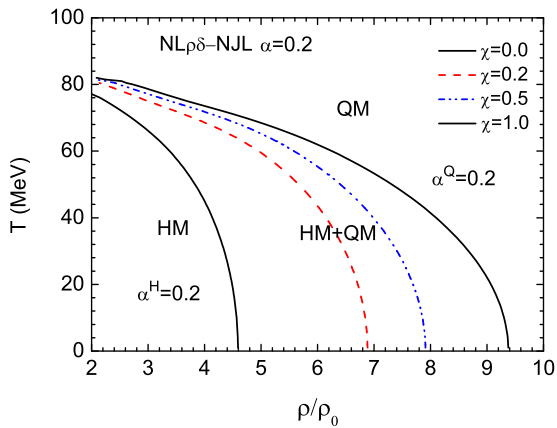}
\caption{\label{fig:nlrhodeltanjl-Trhoasym}(Color on line)~
Phase diagram in the
$T-\rho_B^{}$ plane for asymmetric matter with the parameter set
$NL\rho \delta$.}
\end{center}
\end{figure}

 In Fig.~\ref{fig:nlnjl-Trhosym} we plot the (identical) $T-\rho_B^{}$
 phase diagrams of symmetric
matter. For asymmetric matter  $\alpha=0.2$ in
Fig.~\ref{fig:nlrhonjl-Trhoasym} we show the results for the
parameter set NL$\rho$ and in Fig.~\ref{fig:nlrhodeltanjl-Trhoasym}
for the parameter set $NL\rho \delta$. Now the mixed phase region,
the binodal surface, is rather different. In general the onset
density of the mixed phase in asymmetric matter is smaller than that
in symmetric matter, similar to the results obtained by the MIT-Bag
model~\cite{Muller97, Toro06, Torohq09, Pagliara10, Cavagnoli10}. In
fact the mechanism is the same: for a given baryon density in
isospin asymmetric matter we have a larger repulsion in the hadron
phase since the symmetry term is less important in the quark sector
for both MIT-Bag and NJL effective lagrangians.

If the $NL\rho \delta$ parameter set is used, the onset density will
be further reduced, as shown in
Fig.~\ref{fig:nlrhodeltanjl-Trhoasym}. This is nicely due to the
fact that at high baryon density the symmetry energy of hadron
matter with the parameter set $NL\rho \delta$ is much larger than in
the $NL\rho$ case, see the following.
 All that also indicates that isospin effects can be very important for
hadron-quark phase transition in heavy-ion collision in order to
shed light on nuclear interactions in an hot and dense medium.

In Figs.~\ref{fig:nlnjl-Trhosym},~\ref{fig:nlrhonjl-Trhoasym} and
~\ref{fig:nlrhodeltanjl-Trhoasym} the curves corresponding to a
given quark fraction $\chi$ inside the mixed phase are also
shown. We note that the isospin effects are mainly relevant in the
initial part of the coexistence region. This is consistent with the
interpretation of the pressure behavior in
Fig.~\ref{fig:nlrhonjl-Prhoasym}. We finally remark that such lower
density zone can be reached even at relatively lower beam energies.

To understand the role of the symmetry term on the hadron-quark
phase transition, we further investigate the symmetry energy
$E_{sym}$  in the two phases, defined in general
by~\cite{Toro06,Pagliara10}
\begin{equation}
    (E/A)_{\alpha_i}=(E/A)_{\alpha_i=0}+E_{sym}{\alpha_i}^2
\end{equation}
where $\alpha_i=\alpha^H$ for hadron matter,  $\alpha^Q$ for quark matter.

\begin{figure}[htbp]
\begin{center}
\includegraphics[scale=0.25]{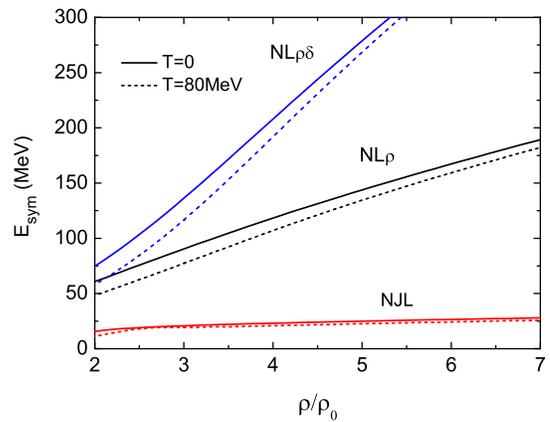}
\caption{\label{fig:esym-T=0-80} (Color on line)~
Symmetry energy of nuclear matter with
the parameter sets
 $NL\rho $, $NL\rho \delta$ and that of $NJL$ quark matter.
  Temperature $T=0,\,80\,MeV$.}
\end{center}
\end{figure}

From the hadron (RMF approach) and the quark (MIT-Bag, NJL) EoSs
described before we can evaluate the symmetry energy as a function
of the baryon density at fixed temperature. We display the result in
Fig.~\ref{fig:esym-T=0-80} for both hadron and quark matter at
T=0, 80\,MeV. It is clear the symmetry energy of quark matter, only
due to the Fermi kinetic term in both MIT and NJL schemes, is much
smaller than that of hadron matter, where also interaction isovector
mesons contribute. Moreover in general we get a decrease of the symmetry 
energy at lower densities and higher temperatures due to a smaller 
contribution from the Fermi kinetic motion.

The symmetry energy of hadron matter with the parameter set $NL\rho
\delta$ is larger than that of the $NL\rho$ case. As discussed
before by comparing Fig.~\ref{fig:nlrhonjl-Trhoasym} and
Fig.~\ref{fig:nlrhodeltanjl-Trhoasym}, this leads to a smaller onset
density of the transition. It is interesting to note that this is a
genuine relativistic effects since the scalar covariant nature of
the isovector $\delta$ meson contributes to increase the symmetry
energy at high densities directly with a larger repulsion and
indirectly via a splitting of the neutron/proton effective masses,
with $M^*_n\,<\,M^*_p$, see details in refs.~\cite{Liu02,Baran05}.

We see that larger is the symmetry energy difference between hadron
and quark phase, smaller is the onset density of the phase
transition. Moreover a big variation of the symmetry term will
strengthen the observable signals of phase transition, which will be
discussed later. Of course all these effects will be enhanced by the
global isospin asymmetry of the system, suggesting the interest in
experiments with very neutron-rich unstable beams.

\begin{figure}[htbp]
\begin{center}
\includegraphics[scale=0.25]{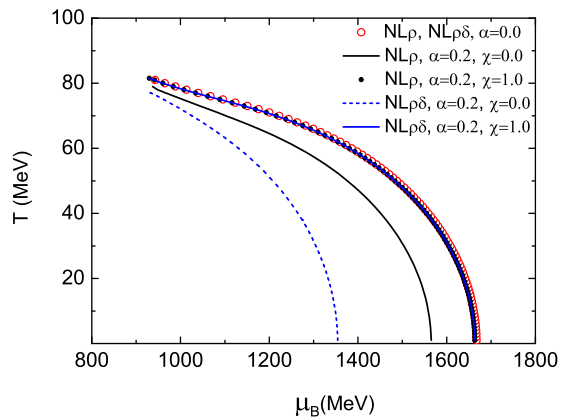}
\caption{\label{fig:T-mu} (Color on line)~
Phase diagram in the $T-\mu_B^{}$ plane for
symmetric and asymmetric matter with the parameter sets $NL\rho$ and
$NL\rho \delta$. Always $NJL$ quark $EoS$.}
\end{center}
\end{figure}

In Fig.~\ref{fig:T-mu} we plot the $T-\mu_B^{}$ phase diagram for
symmetric and asymmetric matter with both the parameter sets
$NL\rho$ and $NL\rho \delta$. For the symmetric matter at a given
temperature,  $\mu_B$ is constant in the mixed phase, i.e. varying
the quark matter fraction $\chi$, then we have only one transition
line (the empty circles). At variance, like in the MIT-Bag
calculation of Fig.~\ref{fig:nlrhomit-Tmuasym}, in the asymmetric
case we have a monotonous $\mu_B$ increase with increasing $\chi$,
so there are two curves in $T-\mu_B^{}$ plane presenting the start
and end, respectively, of the transition from nuclear to quark
matter. What is important is that the chemical potentials of the
onset of the transition ($\chi=0$) in asymmetric matter are always
smaller than the corresponding ones of symmetric matter. This is
more evident with the parameter set $NL\rho \delta$ (dotted curve),
which again shows the importance of symmetry energy effects.

\subsubsection{Isospin Distillation inside the mixed phase}
\begin{figure}[htbp]
\begin{center}
\includegraphics[scale=0.25]{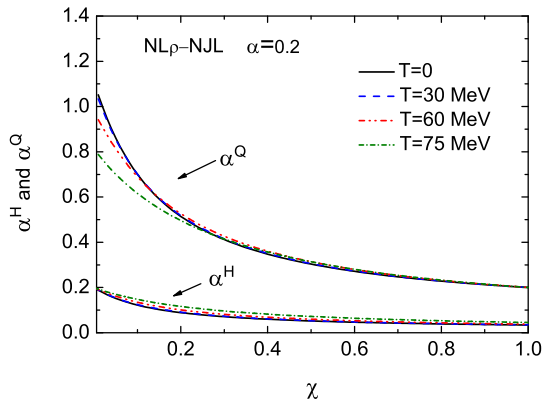}
\caption{\label{fig:nlrhonjl-isodist} (Color on line)~
Asymmetry of quark matter and nuclear 
matter
inside the mixed phase at different temperatures. The parameter set $NL\rho$ 
is used for the hadron phase.}
\end{center}
\end{figure}

\begin{figure}[htbp]
\begin{center}
\includegraphics[scale=0.25]{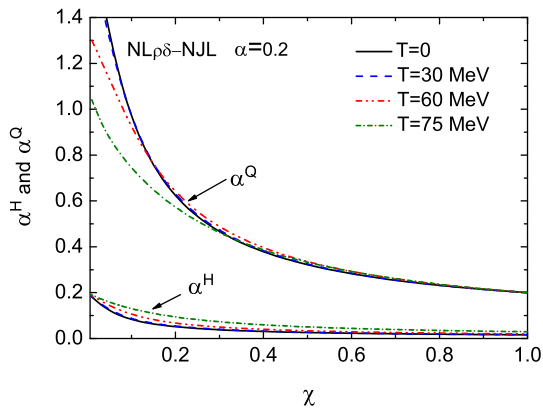}
\caption{\label{fig:nlrhodeltanjl-isodist} (Color on line)~
Like in the previous Figure,
but with the $NL\rho \delta$ $EoS$ for the hadron phase.}
\end{center}
\end{figure}

In Figs.~\ref{fig:nlrhonjl-isodist} and
~\ref{fig:nlrhodeltanjl-isodist}, we show the variation of the
isospin asymmetry parameters of hadron ($\alpha^H$) and quark
($\alpha^Q$) matter inside the mixed phase at various temperatures, with
the global asymmetry $\alpha=0.2$. We clearly see the much larger
values of $\alpha^Q$ when the quark phase starts forming, roughly
for a $\chi$ fraction between 0.0 and 0.4. Of course the effect disappears 
when the pure quark phase is reached ($\chi=1$) where the global asymmetry
is recovered just for the two charges ($\rho_b,~\rho_3$) conservation. 
This is a
nice Isospin Distillation effect ruled by the symmetry energy gap in
the two phases, as confirmed by the enhancement
in Fig.~\ref{fig:nlrhodeltanjl-isodist}, where the more symmetry
repulsive $NL\rho\delta$ $EoS$ is used for the hadron part.

We note that the isospin asymmetry of quark matter decreases with the
enhancement of temperature. This is due to a general decrease of symmetry 
energy effects at higher temperatures (and lower densities), as we can 
clearly see in Fig.~\ref{fig:T-mu}.  

The color pairing interaction at high density can reduce the isospin 
distillation since it is energetically equivalent to the introduction 
of an effective
symmetry repulsion in the quark phase~\cite{Pagliara10,Torohq09}. 
However the isospin effects discussed before are still present. 
Moreover
we can expect the pairing correlations to be less important with increasing 
temperature. 
Experiments focused to observe isospin effects in the mixed phase, 
using neutron-rich heavy ion collisions at intermediate energies, 
appear very appealing.

%

This behavior of the quark isospin asymmetry inside  the mixed phase of 
the hadron-phase transition will affect the following hadronization
in the expansion stage, finally
producing some observable signals in heavy-ion experiments.
As suggested before in Ref.~\cite{Toro06,Torohq09}, an inversion in the trend 
of the emission of neutron rich clusters,
$\pi^-/\pi^+$, $K^{0}/K^{+}$ yield ratios in high density regions, and an 
enhancement of
the production of isospin-rich resonances and subsequent decays may be found.
Besides, there is a controversial point of view about the
enhancement of the yield ratio  
$\bar{\Lambda}/\bar{p}$
~\cite{Stephans97,Armstrong99,Back01,Rapp01,Greiner2001}.

For instance, in~\cite{Toro06} the reaction $^{238}$U+$^{238}$U
~( isospin asymmetry $\alpha=0.227$)
at 1\,$A$~GeV has been investigated in the RMF approach, and a rather exotic 
nuclear matter is
formed with baryon density around $3-4\rho_0$, temperature  $50-60$ MeV, 
likely inside the estimated mixed phase
region, especially with the parameter set $NL\rho\delta$ for 
the hadron sector.

\subsubsection{New effect of the dynamical quark masses: a Critical-End-Point?}

If we compare MIT-Bag (with fixed current quark masses) and NJL 
(with chiral restoration mechanism) results for the mixed phase we remark 
only one main difference. As shown in 
Figs.~\ref{fig:nlrhomit-Trhoasym},
~\ref{fig:nlrhomit-Tmuasym}, phase diagrams in $T-\rho_B^{}$ 
and $T-\mu_B^{}$ planes
are derived, using the MIT Bag model, with a critical 
temperature reached only at zero baryon density. At variance the binodal 
curves at high temperature and
low density cannot be obtained with the NJL model and we see a narrowing of 
the coexistence region up to a kind of $Critical-End-Point$,
 Figs.~\ref{fig:nlnjl-Trhosym},~\ref{fig:nlrhonjl-Trhoasym},
~\ref{fig:nlrhodeltanjl-Trhoasym} and~\ref{fig:T-mu}.
This important result derives from two qualitative new features of the
NJL effective theory: i) the quark masses variation due to the chiral 
restoration, ii) the dependence of the effective Bag-constant $B_{eff}$,
 Eq.~(\ref{Beff}), on temperature and baryon density. 

\begin{figure}[htbp]
\begin{center}
\includegraphics[scale=0.25]{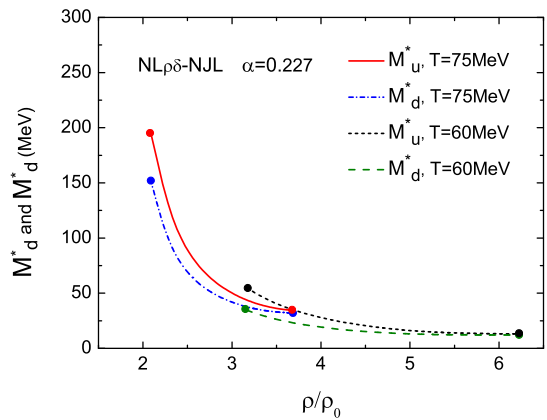}
\caption{\label{fig:quarkmasses} (Color on line)~ 
Dynamical masses of $u,\,d$ quarks inside the mixed phase
at different temperatures. The solid circles indicate the two density 
limits of the
coexistence region. Asymmetric matter with $\alpha=0.227$. The $NL\rho \delta$ 
$EoS$ is used for hadron phase.}
\end{center}
\end{figure}

\begin{figure}[htbp]
\begin{center}
\includegraphics[scale=0.25]{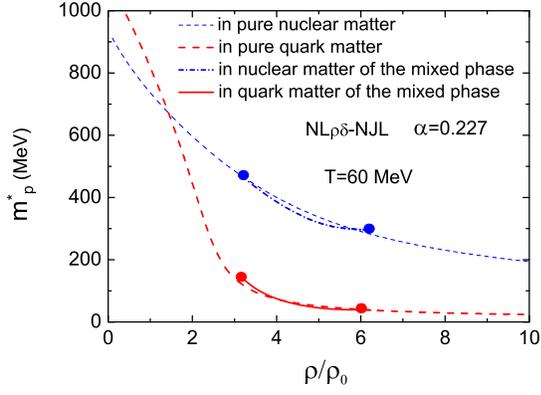}
\caption{\label{fig:protonmass-T=60} (Color on line)~
Effective proton masses in the hadron
(upper) and quark (lower) matter inside the mixed phase at $T=60~MeV$.
The solid circles indicate the two density 
limits of the
coexistence region. Asymmetric matter with $\alpha=0.227$. The $NL\rho \delta$ 
$EoS$ is used for hadron phase.}
\end{center}
\end{figure}

\begin{figure}[htbp]
\begin{center}
\includegraphics[scale=0.25]{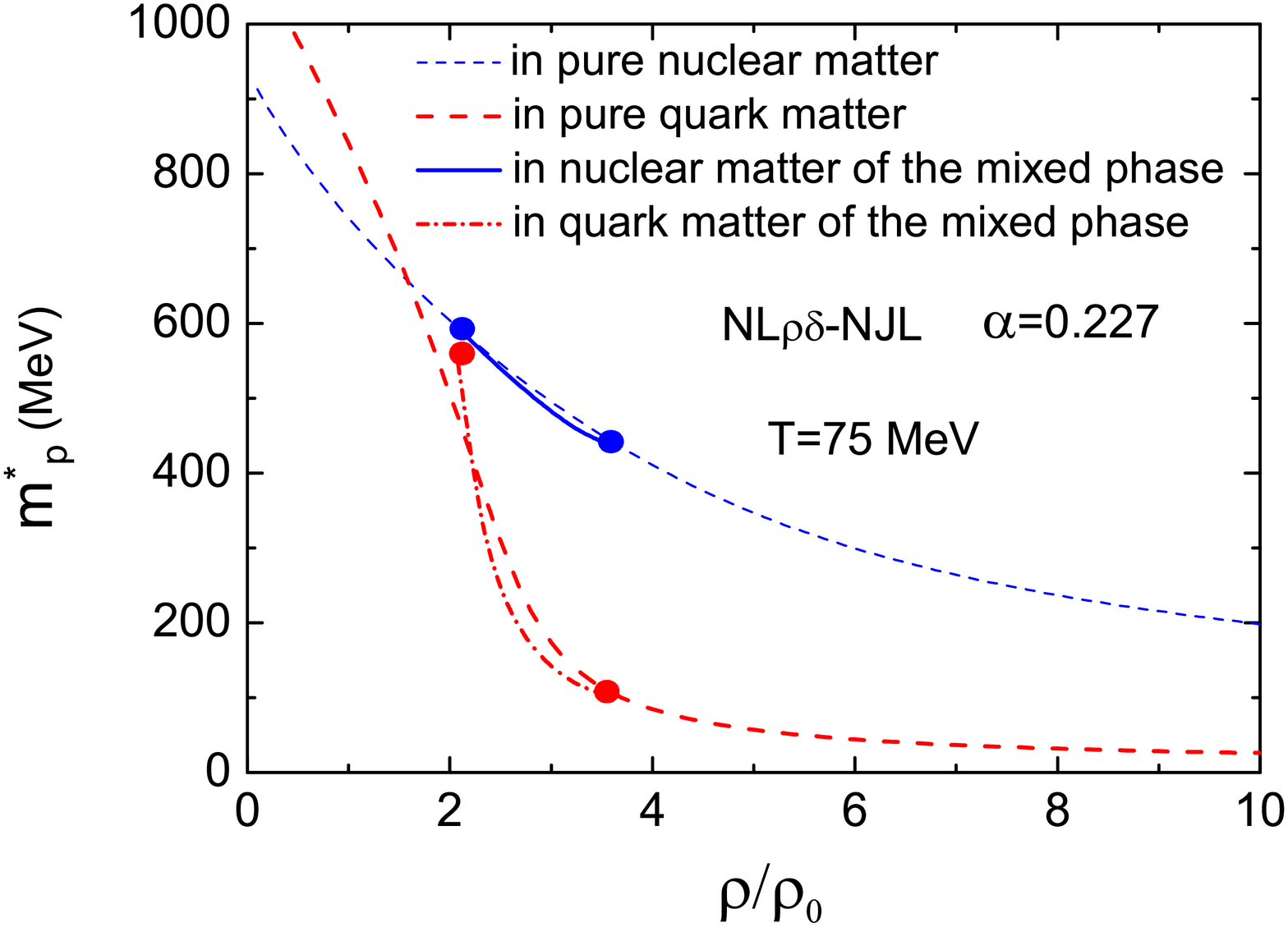}~
\caption{\label{fig:protonmass-T=75} (Color on line)~
Like in the previous Figure,
but at temperature $T=75~MeV$.}
\end{center}
\end{figure}

\begin{figure}[htbp]
\begin{center}
\includegraphics[scale=0.25]{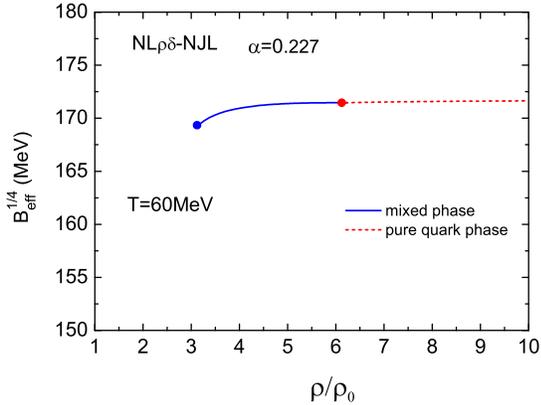}
\caption{\label{fig:Beff-T=60}  (Color on line)~
Baryon density dependence of the NJL Effective
Bag ``constant'', $(B_{eff})^{1/4}$, at $T=60~MeV$.
The line ending with solid circles indicates
coexistence region. Asymmetric matter with $\alpha=0.227$. The $NL\rho \delta$ 
$EoS$ is used for hadron phase.}
\end{center}
\end{figure}

\begin{figure}[htbp]
\begin{center}
\includegraphics[scale=0.25]{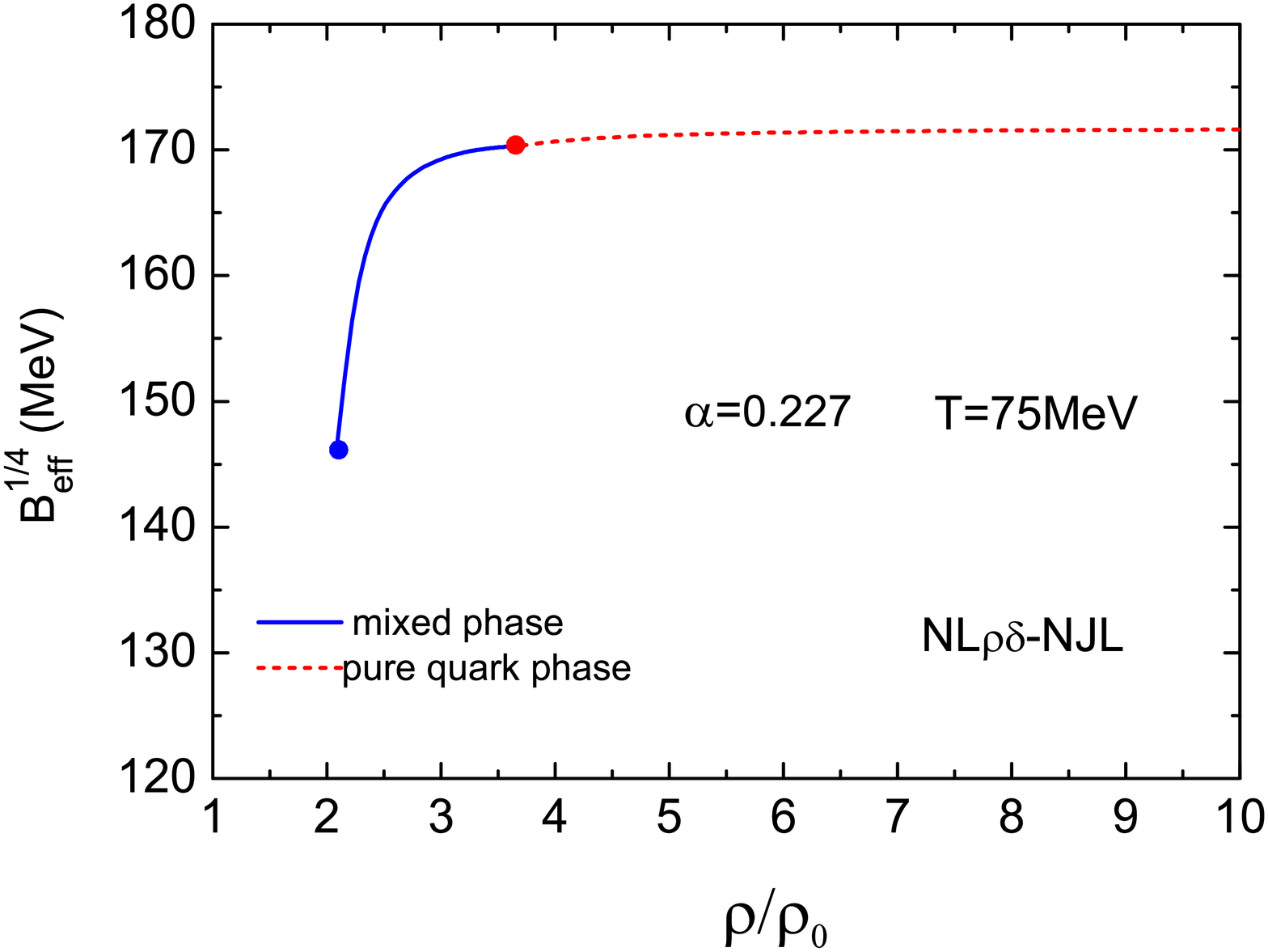}
\caption{\label{fig:Beff-T=75} (Color on line)~
Like in the previous Figure,
but at temperature $T=75~MeV$.}
\end{center}
\end{figure}

The two effects are jointly leading to a determination of an End-Point
of the mixed phase:

\begin{itemize}
\item{
The chiral dynamics largely increases the quark masses at low densities 
and finite temperatures,
see Fig.~\ref{fig:quarkmasses} where we plot the $u, d$ effective masses 
inside the mixed phase at different temperatures, for asymmetric matter.
If we reach the limit of a nucleon effective mass in the hadron phase
smaller than the corresponding combination of quark effective masses,
e.g. for protons $M_p^*\,<\,2M_u^* + M_d^*$, we would have unphysical
solutions for the 
Gibbs conditions since from Eqs.~(\ref{hadrondensity}),~(\ref{quarkdensity})
we will get only $\rho_B^Q\,<\,\rho_B^H$ results.
This is confirmed by the Figs.~\ref{fig:protonmass-T=60},
~\ref{fig:protonmass-T=75}, 
where we plot the proton effective mass in the hadron phase (solid line)
and in the quark phase (dot-dashed line) inside the coexistence region at
$T=60~MeV$ and $T=75~MeV$. We see that at the higher temperature we are close
to a crossing, indication of the lack of physical solution at higher 
temperatures, as already seen in the corresponding $NL\rho\delta$ results
of the previous 
Figs.~\ref{fig:nlrhodeltanjl-Trhoasym} and ~\ref{fig:T-mu}.

We have a final interesting comment about the quark mass splitting shown in 
Fig.~\ref{fig:quarkmasses} for asymmetric matter. The difference in the
quark masses, with $M_d^*\,<\,M_u^*$, is larger at the onset of the mixed
phase, where the isospin distillation effects induces a larger difference in 
the two quark-antiquark  condensates.} 

\item{
Actually we clearly see that in the same $(T,~\rho)$ region we do not have 
solutions of the Gibbs conditions. This is due to the second qualitative 
new feature of the $NJL$ approach, the density and temperature dependence
of the ``Effective Bag Constant'', 
 Eq.~\ref{Beff}, also related to the dynamical quark mass variation. 
As a consequence at low densities and high temperatures, for 
small values of the Bag constant we cannot get mixed phase solutions
since the hadron pressure (mostly thermal) cannot equilibrate
the quark pressure in the coexistence zone. In Figs.~\ref{fig:Beff-T=60},
~\ref{fig:Beff-T=75} we show the density dependence of $(B_{eff})^{1/4}$ 
at $T=60$ and
$T=75~MeV$ (the solid circles give the mixed phase limits). For the $T=75~MeV$
case we see a sudden drop around the onset 
of the mixed phase, good indication of a lack of solution for larger 
temperatures.}
\end{itemize}

In conclusion the chiral dynamics seems to lead to a 
$Critical-End-Point$ ($CEP$) of the first order hadron-quark transition,
around $T\,\simeq\,80~MeV$ and  $\rho_B\,\simeq\,2\rho_0$ or
$\mu_B\,\simeq\,900~MeV$. Beyond this point, i.e. at higher temperatures and
smaller baryon densities, we have to follow an approach with just
one effective $EoS$ able to describe both phases in order to check if we get a 
continous transition. 

It is interesting to note that the $CEP$ appears when the $NJL$ quark masses 
have reached the constituent quark values. 
However we are aware that the NJL approach, with only 
chiral dynamics, is not good at low densities and finite temperatures
since confinement is not accounted for (in fact our $B_{eff}$ goes to zero),
just where we can expect that hadron degrees of freedom will start to play 
a role \cite{Buballa05}. A further investigation including a confinement 
mechanism is certainly required in order to confirm the End-Point 
evaluation and to understand the related physics.
Finally correlation effects, beyond the mean field
approximation, would be also more important.

In any case the fact that in the $(T\,,\,\mu_B)$ plane around the 
End Point we definitely get a derivative $dT/d\mu\,\neq\,0$ 
 (see Fig.~\ref{fig:T-mu}) could be an indication that we are actually reaching
a continous transition for chemical potentials and entropies, as suggested
by the Clausius-Clapeyron Equation \cite{Torohq09}. 

Finally two more remarks:
\begin{itemize}
\item{ From 
 Figs.~\ref{fig:nlnjl-Trhosym},~\ref{fig:nlrhonjl-Trhoasym},
~\ref{fig:nlrhodeltanjl-Trhoasym} and~\ref{fig:T-mu}, we see
that the End-Point of the mixed phase is not depending on the global asymmetry
of the system and on the symmetry term of the used hadron interaction. This is 
consistent with the reduced effect of the symmetry energy at high temperatures 
and low densities, Fig.~\ref{fig:esym-T=0-80}.}
\item{
Isospin effects appear still relevant in the region just below the 
End-Point, T$\subset(50-80)$\,MeV and $\rho_B^{}\subset(2-4)\,\rho_0$,
 accessible in the transient compression stage of Heavy Ion Collisions
at intermediate energies.}
\end{itemize}

The latter point is interesting even because in that phase diagram zone
the NJL and MIT-Bag results are very similar, for similar bag constants
$B_{MIT}\,\simeq\,B_{eff}$ \cite{Torohq09}. This makes more reliable the 
observed isospin effects.

\section{Summary}
In this work, we investigate the hadron-quark phase transition in 
isospin asymmetric matter. We use a Two-EoS approach as in all previous
calculations. The novelty of this study is to insert chiral restoration 
effects on the quark masses using the 3-flavor NJL model for interacting quark
instead of the MIT Bag model. We obtain the binodal surface of a first order
 hadron-quark phase transition in
the region of $\rho_B^{}>2\rho_0$ and temperature $T$ less than 
about 80\,MeV, 
available in heavy-ion experiments in the near future.

The calculated results show that the onset density of the phase transition 
is lowered with
the increasing asymmetry parameter $\alpha$, property possible to be probed 
in the new planned facilities,
for example, FAIR at GSI-Darmstadt and NICA at JINR-Dubna, with realistic 
asymmetries for stable beams, see~\cite{Toro06}.

The phase diagrams at high density are like the ones given by the MIT Bag, 
model with appropriate bag constant,
since the chiral symmetry is restored and the quark masses approach the 
current values.
But the dynamical quark mass becomes more and more important
with the reduction of baryon number density and the increase of temperature,
 causing a reduction of the Effective Bag Constant. The resulting effect
is more relevant just
in the 
region T\,$\subset(50-80)$\,MeV and
$\rho_B^{}\subset(2-4)\,\rho_0$, available with the new planed 
facilities, where some suggested important observables may be found
as the signals of hadron-quark phase transition in the future.

The most interesting effect of the use of a consistent chiral quark mass 
dynamics is the appearance of a kind of Critical-End Point for the first order
transition,
 around $T\,\simeq\,80~MeV$ and  $\rho_B\,\simeq\,2\rho_0$ or
$\mu_B\,\simeq\,900~MeV$. Furthermore the obtained phase diagram
exhibits a region with confinement but chiral restored symmetry,
as expected for the $Quarkyonic$ matter \cite{McLerran10}.

At variance with the results with the MIT-Bag quark $EoS$,
 the phase diagram at lower density and higher temperature
cannot be derived from the Gibbs conditions in a two-EOS model, which means 
that chiral symmetry is very important in the phase transition.
This conclusion stimulates new efforts in the search of an unique $EoS$,
with chiral dynamics and confinement mechanism, able to describe both 
phases in the region just above the suggested Critical-End-Point.

In any case reliable results have been obtained in the interesting region 
T$\subset(50-80)$\,MeV and $\rho_B^{}\subset(2-4)\,\rho_0$,
that will undergo the test of experiments in a near future.

%
%

\bigskip

\begin{acknowledgments}
This  work  was supported in part by the National Natural Science Foundation
of China under Grants Nos. 10875160, 11075037, 10935001 and the Major State 
Basic Research Development Program under Contract No. G2007CB815000.
\end{acknowledgments}


\begin{thebibliography}{150}
\bibitem{hstrange}
Strangeness is not considered in the hadron $EoS$. This is due to two 
main reasons: i) Strange particle production is very limited at the
intermediate energies of interest here; ii) From fully dynamical
calculations of strange particle production it appears that in the
short time interval of the transient high density state the strangeness
chemical equilibrium is not reached \cite{Ferini05,Ferini06}.
\end{thebibliography}
\end{document}